\documentclass[12pt]{article} 
\usepackage{fullpage} 
\usepackage[]{graphicx}

\newcommand{\ket}[1]{|#1 \rangle}
\newcommand{\bra}[1]{\langle#1 |}

\begin{document}

\title{An efficient quantum filter for multiphoton states}

\author{Yuan Liang Lim$^{1}$ and Almut Beige$^{2,1}$ \\[0.3cm]
{\small $^1$Blackett Laboratory, Imperial College London, Prince Consort Road,} \\[-0.1cm]
{\small London SW7 2BZ, United Kingdom} \\[-0.1cm]
{\small $^2$Department of Applied Mathematics and Theoretical Physics, University of Cambridge,} \\[-0.1cm] 
{\small Wilberforce Road, Cambridge CB3 0WA, United Kingdom}}

\date{\today}

\maketitle 

\begin{abstract}
We propose a scheme for implementing a multipartite quantum filter that uses entangled photons as a resource. It is shown that the 
success probability for the 2-photon parity filter can be as high as ${1 \over 2}$, which is the highest that has so far been predicted 
without the help of universal two-qubit quantum gates. Furthermore, the required number of ancilla photons is the least of all current 
parity filter proposals. Remarkably, the quantum filter operates with probability ${1 \over 2}$ even in the $N$-photon case, irregardless 
of the number of photons in the input state. \\[.25cm]
\end{abstract}

\section{Introduction}

Recently, much effort has been made to find efficient schemes for the realisation of non-linear operations between photons, thereby 
aiming at minimising the required resources and increasing success probabilities. This is motivated by the fact that for many practical 
purposes, the polarisation states of photons provide the most favoured qubits since they possess very long lifetimes and an ease in 
distribution. Potential applications include quantum information processing \cite{Gottesman,Knill} as well as quantum cryptography 
\cite{Bennett,Ekert}. However, it is difficult to create an effective interaction between photons. Hence linear optics quantum computing 
requires the presence of auxiliary photon states and postselective measurements which implies, in general, finite success rates of gate 
operations \cite{Sipe}. 

An effectively nonlinear operation, whose implementation has widely been discussed in the literature, is the {\em parity} or {\em quantum 
filter} \cite{Hofmann,Grudka,Zou}. The application of parity filters is diverse, ranging from quantum non demolition measurements of 
entanglement to the generation of multiphoton quantum codes \cite{Hofmann} and the generation of multipartite entanglement \cite{Zou}. 
Moreover, it has been shown that the parity filter can constitute a crucial component for the generation of Cluster states for one-way 
quantum computing \cite{Verstraete,Browne}. Together with single qubit rotations and measurements, the parity filter constitutes a 
universal set of gate operations \cite{Browne}. 

Applied to  two photons, the parity filter projects their state onto a 2-dimensional subspace in which a measurement of the linear 
polarisation (horizontal or vertical) would give the same result for each photon. We denote the corresponding filter operator as $P_2$ 
and define 
\begin{equation} \label{par}
P_2 = \sqrt{p_2} \, \Big( \, \ket{HH}\bra{HH}+\ket{VV}\bra{VV} \, \Big) ~,~
\end{equation} 
where $H$ and $V$ describe a horizontally and a vertically polarised photon, respectively. Besides, $p_2$ is the success probability for 
the performance of the parity projection on an arbitrary input state. This means, even when applied to a parity eigenstate, the photons 
only pass through the filter with probability $p_2$. In this paper, the term {\em success probability} denotes the projection efficiency 
of a given setup.

In the original proposal of a linear optics implementation of the parity filter \cite{Hofmann}, Hofmann and Takeuchi obtained a success 
probability of $p_2={1 \over 16}$ after passing the photons through several beam splitters and performing postselective measurements. Two 
other proposals yield a higher success probability of $p_2 = {1 \over 4}$ \cite{Grudka,Zou}. Grudka and Wojcik achieve this by using the 
idea of teleportation \cite{Knill} and by employing ancilla states containing six photons \cite{Grudka}. Zou and Pahlke use a single mode 
quantum filter that separates the 1-photon state from the vacuum and the 2-photon state. By combining two such single mode filters, a 
parity filter can be realised that requires a 4-photon ancilla state as a resource \cite{Zou}. 

In direct analogy to the 2-photon parity filter (\ref{par}), a quantum filter for $N$ photons can be defined
by the operator
\begin{equation} \label{PN}
P_N=\sqrt{p_N} \, \Big( \, \ket{HH \, . \, . \, . \, H}\bra{HH \, . \, . \, . \, H}+\ket{VV \, . \, . \, . \, V}\bra{VV \, . \, . \, .\, 
V} \, \Big) ~.~
\end{equation} 
Applied to an arbitrary input state with $N$ photons, this filter projects the system with probability $p_N$ onto the 2-dimensional 
subspace where all photons have the same polarisation in the $\ket{H}$ and $\ket{V}$ basis. One way to implement this gate is to pass the 
input state through $(N-1)$ 2-photon parity filters, which succeeds with overall probability $p_N= p_2^{N-1}$. This approach presents a 
steep challenge for large particle numbers, given the above mentioned success probabilities of a single 2-photon parity check.

In the following, we describe a potential implementation of the $N$-photon quantum filter (\ref{PN}) with a success rate as high as $p_N 
= {1\over 2}$, which is much more effective than performing operation (\ref{PN}) with the previously proposed 2-photon parity filters. As 
a resource we require the presence of the $N$-photon GHZ-state
\begin{equation} \label{GHZ}
|A^{(N)} \rangle = {\textstyle {1 \over \sqrt{2}}} \, \Big( \, \ket{HH \, . \, . \, . \, H} + \ket{VV \, . \, . \, . \, V} \, \Big) ~.~ 
\end{equation}
In principle, this photon state can be prepared on demand \cite{Gheri,Lange,Yuan}. Furthermore, we require a photon-number resolving 
detector that can distinguish between 0, 1 and 2 photons. To implement the quantum filter (\ref{PN}) we use ideas that have been inspired 
by a recently performed entanglement purification protocol \cite{Pan} as well as by a scheme for the realisation of quantum logic 
operations between photonic qubits \cite{Franson}.

\section{A multipartite quantum filter} \label{filter}

The most important component of our scheme is the polarising beam splitter, which redirects a photon depending on its polarisation to one 
of the output modes. In the following, $\ket{\lambda_i}$ describes a photon with polarisation $\lambda$ travelling in mode $i$. Besides, 
we denote the input modes $i=1$ and $2$ and the output modes $i=1'$ and $2'$ such that a $V$ polarised photon entering input mode $1$ and 
an $H$ polarised photon entering input mode $2$ leave the setup through output $1'$. Suppose, two photons enter the setup in different 
modes. Then the effect of the beam splitter can then be summarised in the transformation 
\begin{equation} \label{comp}
\ket{\lambda_1 \mu_2} \otimes \ket{0_{1'} 0_{2'}} \longrightarrow 
\left\{ \begin{array}{ll} \ket{0_1 0_2} \otimes \ket{H_{1'}H_{2'}}  ~,~ & {\rm if} ~~ \lambda = \mu = H ~,~ \\[0.1cm]
\ket{0_1 0_2} \otimes \ket{V_{1'}V_{2'}}  ~,~ & {\rm if} ~~ \lambda = \mu =V  ~,~ \\[0.1cm]
\ket{0_1 0_2} \otimes \ket{(HV)_{1'} 0_{2'}}  ~,~ & {\rm if} ~~ \lambda = V ~~ {\rm and} ~~ \mu = H ~, \\[0.1cm] 
\ket{0_1 0_2} \otimes \ket{0_{1'} (HV)_{2'}}  ~,~ & {\rm if} ~~ \lambda = H ~~ {\rm and} ~~ \mu = V ~.~ \end{array} \right.
\end{equation} 
We show now that this operation can be used to realise a filter which compares the polarisation $\lambda$ of a target photon with the 
polarisation of an ancilla photon prepared in $\ket{\mu_2}$. With $\mu$ being either $V$ or $H$, the filter operation corresponds to the 
projector $\ket{\mu} \bra{\mu}$ and can be implemented with {\em unit} efficiency.

Suppose a photon number resolving detector is placed in one of the output modes, say output $2'$, and the target photon enters the system 
prepared in $\ket {\lambda_1} =\alpha \, \ket{H_1} + \beta \, \ket{V_1}$. Then the target photon is found to have polarisation $\mu$ in 
case of a single click in the detector indicating a 1-photon measurement. Using Eq.~(\ref{comp}), one can calculate the corresponding 
unnormalised output state. It is either $\alpha \, \ket{H_{1'}}$ or $\beta \, \ket{V_{1'}}$, depending on whether $\mu$ equals $H$ or 
$V$. Note that the probability for a 1-photon detection ($|\alpha|^2$ or $|\beta|^2$, respectively) is exactly what one would expect 
after applying the filter operation $\ket{\mu} \bra{\mu}$ with efficiency 1 to the incoming photon. Remarkably, the target photon is 
effectively not destroyed in the process. The reason is that it does not matter whether the detector absorbs the target photon or the 
ancilla photon, if both have the same polarisation and are anyway indistinguishable.
 
\subsection{The 2-photon case} \label{two}

Let us now describe how the polarising beam splitter (\ref{comp}) can be used for the implementation of a 2-photon parity filter. The 
setup we consider here contains two polarising beam splitters and two polarisation sensitive detectors (see Figure 1). The target state 
enters the setup via the input modes $1$ and $3$. We further require the presence of the 2-photon ancilla state $\ket{A^{(2)}}$, which is 
a 2-photon Bell state. The ancilla photons should enter the setup via the input modes $2$ and $4$. The two detectors are placed in the 
output modes $2'$ and $4'$. In case the two photons are found in a parity eigenstate, they leave the setup via the output modes $1'$ and 
$3'$ .

\begin{figure}
\begin{center}
\begin{tabular}{c}
\includegraphics[height=4.5cm]{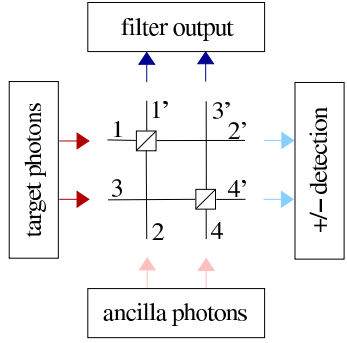}  \\[-0.2cm]
\end{tabular}
\end{center}
\caption[example]{\small Experimental setup for the realisation of a 2-photon parity filter. The target photons enter the setup via the 
input modes 1 and 3, while the ancilla photons enter the setup via inputs 2 and 4. Within the setup, each photon has to pass one 
polarising beam splitter. Under the condition of the detection of one photon in each of the outputs $2'$ and $4'$, the filter succeeded 
and the projected output state leaves the system via the modes $1'$ and $3'$.} \label{fig1}  
\end{figure}  

In the following, we consider the general input state 
\begin{equation}
\ket{\psi_{\rm in}^{(2)}} = \alpha \, \ket{H_1 H_3} +\beta \, \ket{V_1 V_3} +\gamma \, \ket{H_1 V_3} + \delta \, \ket{V_1 H_3}  ~.~
\end{equation}
Our aim is to eliminate the components, where the photons are of different polarisation. Together with the ancilla state $|A^{(2)} 
\rangle$, the setup in Figure \ref{fig1} is entered by the 4-photon state
\begin{eqnarray} \label{in}
\ket{\tilde \psi_{\rm in}^{(2)}} &=& \ket{\psi_{\rm in}^{(2)}} \otimes \ket{A^{(2)}} \nonumber \\
&=& {\textstyle {1 \over \sqrt{2}}} \, \Big( \, \alpha \, \ket{H_1H_2H_3H_4} +  \alpha \, \ket{H_1V_2H_3V_4} + \beta \, 
\ket{V_1V_2V_3V_4} +  \beta \, \ket{V_1H_2V_3H_4} \nonumber \\
&& + \gamma \, \ket{H_1H_2V_3H_4} +  \gamma \, \ket{H_1V_2V_3V_4} + \delta \, \ket{V_1V_2H_3V_4} +  \delta \, \ket{V_1H_2H_3H_4} \, \Big) 
~.~
\end{eqnarray}
We now show that the system can act like a parity filter, if one photon is collected in output mode $2'$ and another one is collected in 
output mode $4'$. Using Eq.~(\ref{comp}), one can show that the 4-photon state (\ref{in}) becomes in this case the unnormalised state
\begin{equation} \label{bus}
\ket{\tilde \psi_{\rm out}^{(2)}} = {\textstyle {1 \over \sqrt{2}}} \, \Big( \, \alpha \, \ket{H_{1'}H_{2'}H_{3'}H_{4'}} + 
\beta \, \ket{V_{1'}V_{2'}V_{3'}V_{4'}} \, \Big) ~.~
\end{equation} 
We further assume that the detectors measure the polarisation of the incoming photons in the rotated basis defined by the 1-photon states 
\begin{equation} \label{tractor}
\ket{\pm} \equiv {\textstyle {1 \over \sqrt{2}}} \, \Big( \, \ket{H} \pm \ket{V} \, \Big) ~.~ 
\end{equation}
It is important that the detectors distinguish the polarisation of each incoming photon in this basis (opposed to just absorbing the 
photon), since this approach guarantees that the output becomes the expected pure state. Using the definition (\ref{tractor}), we can 
rewrite the state (\ref{bus}) as 
\begin{eqnarray} 
\ket{\tilde \psi_{\rm out}^{(2)}} &=& {\textstyle {1 \over \sqrt{2}}} \, \Big( \, \alpha \, \ket{H_{1'}H_{3'}} + \beta \, 
\ket{V_{1'}V_{3'}} \, \Big) \otimes {\textstyle {1 \over 2}} \, \Big( \, \ket{+_{2'}+_{4'}} + \ket{-_{2'}-_{4'}} \, \Big) \nonumber \\
&& + {\textstyle {1 \over \sqrt{2}}} \, \Big( \, \alpha \, \ket{H_{1'}H_{3'}} - \beta \, \ket{V_{1'}V_{3'}} \, \Big) \otimes {\textstyle 
{1 \over 2}} \, \Big( \, \ket{+_{2'}-_{4'}} + \ket{-_{2'}+_{4'}} \, \Big) ~.~
\end{eqnarray} 
Suppose the photons in output ports $2'$ and $4'$ are absorbed in the measurement process. Then the output state of the system equals in 
case of a single click in each of the detectors
\begin{equation}
\ket{\psi_{\rm out}^{(2)}}  = {\textstyle {1 \over \sqrt{2}}} \, \Big( \, \alpha \, \ket{H_{1'} H_{3'}} \pm \beta \, \ket{V_{1'} V_{3'}} 
\, \Big) ~.~
\end{equation} 
The ``$+$" sign applies when both detectors measure the same polarisation (which happens with probability ${1 \over 2}$); the ``$-$" sign 
applies when both detectors measure different polarisations (which also happens with probability ${1\over 2}$). The implementation of the 
parity filter now requires only one further operation, namely the correction of the sign of the state $\ket{V_{1'} V_{3'}}$ in case of a 
sign error. This can be implemented with the help of a Pauli $\sigma_z$ operation on one of the photons and yields the final state
\begin{equation} \label{car}
\ket{\psi_{\rm out}^{(2)}} = {\textstyle {1 \over \sqrt{2}}} \Big( \, \alpha \, \ket{H_{1'} H_{3'}} + \beta \, \ket{V_{1'} V_{3'}} \, 
\Big) ~.~
\end{equation} 
A closer look at the normalisation of this state tells us that the parity filter shown in Figure \ref{fig1} works with efficiency $p_2 = 
{1 \over 2}$. If the success probability of the scheme would be 1, the output state (\ref{car}) would be $\alpha \, \ket{H_{1'} H_{3'}} + 
\beta \, \ket{V_{1'} V_{3'}}$. It can be shown that the filter can also be operated with mixed states as inputs.

\subsection{The $N$-photon case} \label{more}

\begin{figure}
\begin{center}
\begin{tabular}{c}
\includegraphics[height=6.2cm]{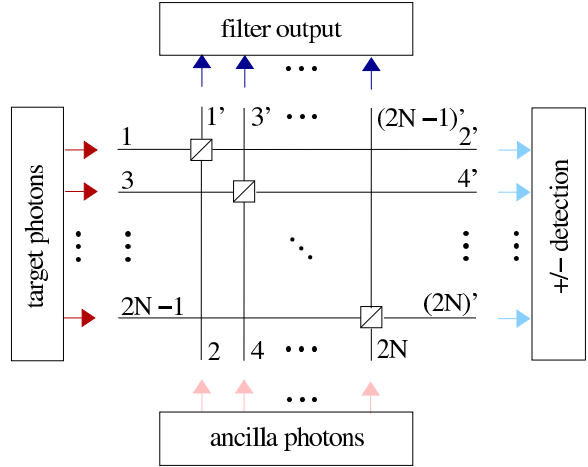} \\[-0.2cm]
\end{tabular}
\end{center}
\caption[example]{\small Experimental setup for the realisation of the $N$-photon quantum filter (\ref{PN}). The $N$ polarising beam 
splitters each compare the state of one of the target photons with the state of one of the ancilla photons, which are initially prepared 
in the GHZ state $|A^{(N)} \rangle$. Besides, $N$ detectors perform photon measurements in the polarisation basis (\ref{tractor}). The 
output photons leave the system via the odd numbered output ports.} \label{fig2}  
\end{figure}  

The generalisation of the above described 2-photon parity filter to the $N$-photon quantum filter (\ref{comp}) is straightforward and 
requires $N$ polarising beam splitters and $N$ polarisation sensitive detectors (see Figure \ref{fig2}). One side of the setup is entered 
by the $N$-photon input state $|\psi_{\rm in}^{(N)} \rangle$ with one photon in each odd-numbered input mode, while the other side is 
entered by an $N$-photon ancilla state $|A^{(N)} \rangle$ with one photon in each even numbered input mode. In the following we denote 
the modes containing the detectors by $2'$, $4'$, ..., $(2N)'$, while the modes $1'$, $3'$, ..., $(2N-1)'$ contain the output state.

Again, the successful operation of the quantum filter is indicated by a single click in each of the detectors. Suppose $\alpha$ denotes 
the amplitude of the state $|H_1H_3 \, . \, . \, . \, H_{2N-1} \rangle$ while $\beta$ is the amplitude of the state $\ket{V_1V_3 \, . \, 
. \, . \, V_{2N-1}}$ with respect to the target state $\ket{\psi_{\rm in}^{(N)}}$. Then we find, using Eq.~(\ref{comp}) and in analogy to 
Eq.~(\ref{bus}), that the collection of one photon in each of the detector output ports transforms the total input state $\ket{\tilde 
\psi_{\rm in}^{(N)}} = \ket{\psi_{\rm in}^{(N)}} \otimes \ket{A^{(N)}}$ into 
\begin{equation} \label{bicycle}
\ket{\tilde \psi_{\rm out}^{(N)}} = {\textstyle {1 \over \sqrt{2}}} \, \Big( \, \alpha \, \ket{H_{1'}H_{2'} \, . \, . \, . \, H_{(2N)'}} 
+ \beta \, \ket{V_{1'}V_{2'} \, . \, . \, . \, V_{(2N)'}} \, \Big) ~.~
\end{equation} 
For the same reason as in the 2-photon case, we assume that the detectors measure the polarisation of the incoming photons in the 
polarisation basis (\ref{tractor}) by absorption. Suppose $J$ is the number of photons found in the $\ket{-}$ state, then one can show 
using Eq.~(\ref{bicycle}) and proceeding as in Section \ref{two} that the output state of the remaining $N$ photons equals 
\begin{equation}
\ket{\psi_{\rm out}^{(N)}} = {\textstyle {1 \over \sqrt{2}}} \, \Big( \, \alpha \, \ket{H_{1'}H_{3'} \, . \, . \, . \, H_{(2N-1)'}} + 
(-1)^{J} \, \beta \, \ket{V_{1'}V_{3'} \, . \, . \, . \, V_{(2N-1)'}} \, \Big) ~.~
\end{equation} 
Depending on whether $J$ is even or odd, the implementation of the quantum filter (\ref{PN}) might further require a conditional sign 
flip on one of the photons, which then yields the final state  
\begin{equation} \label{tram}
\ket{\psi_{\rm out}^{(N)}} = {\textstyle {1 \over \sqrt{2}}} \, \Big( \, \alpha \, \ket{H_{1'}H_{3'}...H_{(2N-1)'}} + \beta \, 
\ket{V_{1'}V_{3'}...V_{(2N-1)'}} \, \Big) ~.~
\end{equation} 
This is exactly the output state that one expects after the application of the quantum filter (\ref{PN}) to the input state $|\psi_{\rm 
in}^{(N)} \rangle$ with success probability $p_N = {1 \over 2}$, which is the highest that has been predicted so far without the use of 
universal two-qubit quantum gate operations.\footnote{A straightforward way of implementing the quantum filter (\ref{PN}) with universal 
two-qubit quantum gate operations (which are difficult to realise with linear optics alone) is to replace each polarising beam splitter 
in the setup (see Figure \ref{fig2}) by a controlled-NOT gate. Furthermore the detectors in all even numbered output modes should perform 
a polarisation sensitive measurement in the $H/V$ basis. The projection efficiency of such a scheme would only be limited by the success 
probability $p$ of a single controlled-NOT operation and would scale like $p^N$. For sufficiently large photon numbers $N$, this might 
decrease below ${1 \over 2}$. Therefore the use of polarising beam splitters, which can operate with a very high fidelity, should be 
favoured \cite{Pan}.} 
Naively, one might expect that the efficiency of the filter decreases with the number of photons in the setup but this is not the case. 
Only the number of photons, which form the ancilla state (\ref{GHZ}), depends on $N$. Furthermore, we remark that the described quantum 
filter also works for mixed $N$-photon input states. 

\subsection{The effect of imperfect photon sources and detectors}

Finally, we briefly discuss the effect of potential imperfections of the experimental setup on the performance of the proposed quantum 
filter operation. The main source for errors, which diminishes both, the success probability and fidelity of the filter output, are the 
use of imperfect ancilla states ({\ref{GHZ}) and the use of photon detectors with finite detector efficiency. Since the setup shown in 
Figure \ref{fig2} is relatively simple, errors arising from faulty alignments of optical elements can be minimised easily. Note that for 
the scheme to work, it is important that each target photon and the corresponding ancilla photon overlap within their coherence time 
within the polarising beam splitter. 

Suppose the ancilla photons do not enter the setup in the GHZ state (\ref{GHZ}) and their state possesses a small component with 
amplitude $\epsilon$ of a state in which the photons are not all of the same polarisation, say $\ket{H_{2}V_{4}V_{6}...V_{2N}}$. In this 
case, the component $\gamma \, \ket{H_{1}V_{3}V_{5}...V_{(2N-1)}}$ of the photon target state also contributes to the case of the 
detection of one photon per output port. Proceeding as in Section \ref{more}, one can calculate the state, in which the photons leave the 
setup in this case. Doing so we find that the fidelity of the filtered output state decreases, in first order in $|\epsilon|^2$, from 
being one to $1-|\epsilon \gamma|^2/ \big( |\alpha|^2 + |\beta|^2 \big)$. Moreover, the probability for the photons to pass the filter 
now differs  by the amount $|\epsilon|^2 \, \big( |\gamma|^2- {1 \over 2} |\alpha|^2 - {1 \over2} |\beta|^2 \big)$ from the desired 
probability ${1 \over 2} \big( |\alpha|^2 + |\beta|^2 \big)$. However, these deviations from the ideal quantum filter are negligible as 
long as $|\epsilon|^2 \ll 1$. We believe that preparation of the ancilla states considered in this paper with a high fidelity and success 
rate (i.e.~with $|\epsilon|^2 \ll 1$) will be feasible in the nearer future \cite{Gheri,Lange,Yuan}. 

Let us now consider a setup, where the photon detector efficiency is not given by $\eta =1$ but by $\eta < 1$. Moreover, let us assume a 
perfectly prepared ancilla state and a photon target state of the special form $|\psi_{\rm in}^{(N)} \rangle = \alpha \, 
\ket{H_{1}H_{3}...H_{(2N-1)}} + \beta \, \ket{V_{1}V_{3}...V_{(2N-1)}}$. For this situation one can easily show that the imperfection of 
the photon detectors reduces the success rate of the filter operation from ${1 \over 2}$ to ${1 \over 2} \eta^N$. In general, for 
arbitrary photon target states, the fidelity of the filter operation is reduced and the success rate of the proposed quantum filter can 
become very small in case of large photon numbers $N$ and $\eta < 1$. Further reductions  of the filter fidelity and success rate can 
arise from the presence of dark counts. Such errors also play a crucial role in previously proposed implementations of quantum filter 
operations \cite{Hofmann,Grudka,Zou}. However, the scheme we propose here has a significantly higher success rate in the ideal case and 
is experimentally less demanding than what has been reported in Refs.~\cite{Hofmann,Grudka,Zou}.

\section{Conclusions}

We described the realisation of a 2-photon parity filter that requires only two polarising beam splitters, two photons prepared in a 
maximally entangled Bell state and two polarisation sensitive detectors. In the presence of ideal photon sources and photon detectors the 
success rate of the scheme equals $p_2 = {1 \over 2}$ and is the highest that has been predicted so far without the help of universal 
two-qubit quantum gate operations and is reached here due to employing an entangled ancilla state as a resource. A generalisation of the 
proposed scheme to the $N$-photon case is straightforward. We showed that the quantum filter (\ref{PN}) can be implemented with the help 
of $N$ polarising beam splitters and an $N$-photon GHZ state as a resource. Remarkably, the success rate of the filter remains ${1 \over 
2}$, irregardless of the size of the input state.

To implement the quantum filter (\ref{PN}), the $N$ polarising beam splitters compare the state of the incoming photons pairwise with the 
state of the ancilla photons. In Section \ref{filter}, we showed that a single polarising beam splitter can be used to realise a filter, 
which measures polarisation $H$ or $V$, respectively, with unit efficiency. Preparing the ancilla photons, for example, in the state 
$\ket{HH \, . \, . \, . \, H}$, would result in a filter that measures whether all target photons are prepared in $\ket{H}$. However, 
since we compare the input state with a GHZ state, which contains two terms, namely $\ket{HH \, . \, . \, . \, H}$ and $\ket{VV \, . \, . 
\, . \, V}$, the probability of the described filter is only as high as ${1 \over2}$. Indeed, the highly entangled $N$-photon GHZ state 
acts as a ``mask" for the filter. \\[0.1cm]

\noindent {\em Acknowledgement.}  
The authors thank  T. Rudolph and F. Verstraete for stimulating discussions.  Y. L. L. acknowledges the DSO National Laboratories in 
Singapore for funding as a PhD student and A. B. thanks the Royal Society and the GCHQ for funding as a James Ellis University Research 
Fellow. This work was supported in part by the European Union and the UK Engineering and Physical Sciences Research Council.

\end{document}